\documentclass{jetpl}
\usepackage[koi8-r]{inputenc}
\usepackage[russian]{babel}
\usepackage{color,amsmath,amsfonts,graphics,epsfig,amssymb}

\twocolumn

\lat

\title{Axion-like particles and the propagation of gamma rays over
astronomical distances
}

\rtitle{Axion-like particles and gamma rays}

\sodtitle{Axion-like particles and the propagation of gamma rays over
astronomical distances}

\author{S.\,V.\,Troitsky\thanks{e-mail: st@ms2.inr.ac.ru}}

\rauthor{S.\,V.\,Troitsky}

\sodauthor{S.\,V.\,Troitsky}

\address{Institute for Nuclear Research of the Russian Academy of
Sciences,\\
60th October Anniversary prospect 7A, 117312 Moscow, Russia}

\dates{October 31, 2016}{*}

\abstract{
In this mini-review, possible manifestations of mixing between axion-like
particles (ALPs) and energetic photons propagating over astronomical
distances are considered. We discuss the evidence for the anomalous
transparency of the Universe from observations of ensembles of distant
gamma-ray sources, present the general formalism for the ALP-photon mixing
and explain how this mechanism may remove the anomaly. We present
relevant values of ALP parameters and discuss future ways to verify the
scenario and to discover the particle in question.}

\begin{document}

\maketitle
\section{Introduction}
\label{sec:intro}
The axion \cite{PQ, a1, a2} is a hypothetical pseudoscalar particle
coupled to
gluons, originally invoked to solve the strong CP problem. Its
characteristic feature is a two-photon coupling which is used to search
for axions both in laboratory experiments and in astrophysical
observations. In various axion models, the mass of the pseudoscalar $m$ is
related to the two-photon coupling $g_{a\gamma \gamma }$ as
\begin{equation}
\frac{g_{a\gamma \gamma }}{10^{-10}~{\rm GeV}}= C
\frac{m}{1~{\rm eV}},
\label{Eq:*}
\end{equation}
where $C$ is a constant of order one. It is a nontrivial task to solve the
strong CP problem if the condition (\ref{Eq:*}) is not satisfied (see
however Refs.~\cite{9703409, 0009290, 0409185}), but light pseudoscalars
which do not obey the relation (\ref{Eq:*}) arise in numerous extensions
of the Standard Model of particle physics (see e.g.\ Ref.~\cite{Ringw-rev}
for a review). These particles are called axion-like particles (ALPs).
Depending on their interactions, they may constitute the dark matter, or a
part of it \cite{ALP-DM}. As we will see below, the existence of a
particle of this kind with certain parameters is actually favoured by
recent astronomical observations.

\section{Anomalous transparency of the
Universe}
\label{sec:transparency}
Thanks to the ALP-photon interaction, conversion of a photon to ALP and
back may happen in the external magnetic field (see Sec.~\ref{sec:mixing}
for a quantitative description). Since ALP interactions are very weak,
this particle penetrates the media which is non-transparent for photons.
This mechanism allows for
the classical ``light shining through the wall'' experiment, in which
photons are supposed to convert to axions or ALPs and back on either side
of a nontransparent wall. Here we note that this experiment is continuosly
being repeated at the scale of the Universe. Indeed, the Universe is
filled with diffuse radiation of various frequencies. Energetic photons
scatter on the soft background radiation when propagating through the
Universe by producing electron-positron pairs \cite{pair-prod}. For gamma
rays with energies between 100~GeV and a few TeV, the principal target is
the infrared background. The density of background infrared photons is
poorly known experimentally but can be constrained from below by summing
the observed light from galaxies \cite{EBLlower-bounds}. The mean free
path of $\sim$TeV gamma rays with respect to the pair production does not
exceed dozens of Megaparsecs. However, a lot of more distant TeV sources
have been detected \cite{detected}. While, in each particular case, this
might be explained by unusual hardening of the emitted spectrum at high
energies, no working mechanism resulting in this hardening is known
\cite{hardening}. Moreover, this interpretation is not supported by the
analysis of the ensemble of all observed sources. Even for the most
conservative models of the infrared background, strength and positions of
upward breaks in the emitted spectra of distant blazars are
redshift-dependent, indicating incorrect account of the absorption
(anomalous transparency of the Universe).

Indeed, atmospheric Cerenkov telescopes, as well as Fermi LAT, continue to
discover very-high-energy (VHE, energies larger than 100~GeV) gamma
radiation from distant sources \cite{hardening, 3c279, 2photons}.
Additional suppression of the VHE flux from distant sources with respect
to that of similar astrophysical objects located close to the observer is
expected and, indeed, has been observed in the Fermi-LAT data
\cite{Fermi-suppr}. It is, unfortunately, hardly possible to compare the
observed suppression with the one expected theoretically. The reason is
that the relevant EBL density is poorly known (see e.g.\
Refs.~\cite{HDwek, Costamante} for reviews). To measure the extragalactic
infrared background while staying within the Solar system is a
challenge because of the overwhelming contribution of the Zodiacal light.
Existing theoretical models of the EBL give different predictions.
However, the lowest possible amount of intergalactic infrared light is
constrained \cite{EBLlower-bounds} by simple counting of contributions
from observed galaxies.

The distant gamma-ray sources under discussion are blazars,
that is active galactic nuclei with relativistic jets pointing
to the observer.
The mechanism of high-energy emission of
blazars is still not figured out definitely, but
their spectral energy distributions are well studied observationally.
For nearby sources, they consist of two wide bumps: the low-energy
one is due to the synchrotron radiation of relativistic electrons
while the high-energy one is
probably related to the inverse Compton scattering.
In the frameworks of a particular EBL model, it is possible
to reconstruct the emitted gamma-ray spectrum of the source by correcting
the observed spectrum for the pair-production suppression.
These ``deabsorbed'' spectra of distant sources
often exhibit hardening, or upward breaks,
not seen in spectra of relatively nearby blazars.

For individual sources, spectral hardenings at high energies have been
found in many cases (see e.g.\ Ref.~\cite{hardening}). But the most
serious arguments in favour of the anomalous transparency of the Universe
come from studies of \textit{ensembles} of distant sources.
Indeed, it was found that the energies corresponding to
upward breaks in deabsorbed spectra of blazars change with redshift and
always correspond to the energy at which the absorption
becomes important (Ref.~\cite{HornsMeyer}, a sample of 7 blazars with
redshifts $z\lesssim 0.536$ observed by imaging atmospheric Cerenkov
telescopes (IACTs) at optical depths $\tau>2$ with respect to the pair
production). Moreover, the strength of the break, that is the difference
between the power-law spectral indices below and above the break point,
also changes with the redshift and does not depend on the properties of a
particular blazar (Ref.~\cite{gamma}, a sample of 20
blazars at optical depths $\tau>1$, IACTs and FERMI LAT, $z\lesssim
2.156$). These effects are surprising because physically, closeby and
distant blazars are very similar.

Statistical significance of these results is determined by the probability
that a similar or stronger effect may be obtained from a fluctuation in a
random data set. This probability is often expressed in terms of standard
deviations $\sigma$, which however have straightforward interpretation for
the Gaussian distribution only. Stated in this way, the significance of
the redshift dependence of the break position \cite{HornsMeyer} is
4.2$\sigma$, while that of the redshift dependence of the break strength
\cite{gamma} is 12.4$\sigma$.
These significance estimates are based on statistical analyses
only while the results may be subject to systematic uncertainties. For
instance,
Ref.~\cite{HornsMeyer} demonstrates that, under the worst assumptions
about systematic errors, the quoted significance of the effect is reduced
by $\sim 1.6 \sigma$.
However, a detailed quantitative study of
systematic uncertainties cannot be performed without a complete sample of
sources.
Indeed, IACTs have narrow field of view and the choice of objects to be
observed is determined by humans, not derived from a uniform sample.

A solution to the problem of the unphysical spectral hardenings requires a
reduction of the gamma-ray attenuation
by means of some mechanism.
But the physics behind the usual
deabsorption procedure is standard and the assumptions about the photon
background are very conservative.
One concludes that only new physics or astrophysics may explain the effect.

The only astrophysical explanation was suggested in Ref.~\cite{Kusenko}.
It assumes that the very same gamma-ray blazars produce also a sufficient
amount of ultra-high-energy cosmic protons.
Interactions of the protons on their way from the source to the Earth
result in secondary photons which need a shorter way to reach the
observer. Protons are charged, and their trajectories are bend in magnetic
fields.
Unless extragalactic magnetic fields are
as low as $\lesssim 10^{-17}$~G everywhere along the line of sight
(including potentially crossed clusters and filaments), this scenario may
have tensions with the observation of VHE variability of 4C$+21.35$ at
the timescale of hours \cite{PKS1222+21}.

\section{ALP explanations}
\label{sec:mixing}
The best-elaborated new-physics explanation involves ALPs (another
possibility is to assume violation of the Lorentz invariance which may
affect the pair-production cross section).
An ALP should mix with photons in external magnetic
fields~\cite{sikivie, raffelt}, and this mechanism allows to suppress the
attenuation due to pair production. ALPs do not produce pairs, and
gamma-ray photons may convert to ALPs,
then travel without attenuation and at some moment convert back to photons.
The photon beam is attenuated, but the ALP beam is not and the overall flux
suppression is less severe.

For estimates of the probability of conversion in various astrophysical
environments, we follow
Ref.~\cite{FRT}.
The notations we use are determined by the following ALP-photon Lagrangian,
\[
\mathcal{L}=\frac{1}{2}(\partial^\mu a\partial_\mu a-m^2 a^2)
-\frac{1}{4}\frac{a}{M}F_{\mu\nu}\widetilde
F^{\mu\nu}-\frac{1}{4}F_{\mu\nu}F^{\mu\nu},
\]
where $F_{\mu\nu}$ is the electromagnetic stress tensor and
$\widetilde
F_{\mu\nu}=(1/2)\epsilon _{\mu \nu \rho \lambda }F_{\rho \lambda }$
is its dual and
$a$ denotes the ALP.
The photon/ALP mixing in the magnetic field
is determined by  the $F_{\mu\nu}\tilde{F}^{\mu\nu}$ term (photon
components
with different polarizations mix also); $g_{a\gamma \gamma } \equiv 1/M$.

Suppose that photons propagate through a region of constant magnetic
field. Then the probability to detect an ALP at the distance $L$ for the
pure-photon initial beam is
\begin{equation}
P=
\frac{4 \Delta _M^2}{\left(\Delta _p+\Delta_{Q,\perp}-\Delta_m
\right)^2+4 \Delta_M^2 }
\sin^2\left( \frac{1}{2}{L \Delta_{\rm osc}}
\right),
\label{eq:conversion}
\end{equation}
where
\[
\Delta_{\rm osc}^2=\left(\Delta_p+\Delta_{Q,\perp}-\Delta_m \right)^2+4
\Delta_M^2
\]
and
we used the following notations,
\[
\begin{array}{rcccl}
\displaystyle \Delta_{M\! i} \!\!\!&=&\!\!\! \displaystyle \frac{B}{2M}
\!\!\!&=&\!\!\! \displaystyle
153
\left(\frac{B}{1~\mbox{G}}\right)
\left(\frac{10^{10}~\mbox{GeV}}{M}\right)\mbox{pc}^{-1}\!,
\\
%\]
%\[
\displaystyle \Delta_m \!\!\!&=&\!\!\! \displaystyle \frac{m^2}{2
\omega} \!\!\!&=&\!\!\! \displaystyle 7.8 \! \times
\! 10^{-4}\left(\frac{m}{10^{-7}~\mbox{eV}}\right)^2
\! \left(\frac{1~\mbox{TeV}}{\omega}\right)\mbox{pc}^{-1}\!,
\\
%\]
%\[
\displaystyle \Delta_p \!\!\!&=&\!\!\! \displaystyle \frac{\omega
_p^2}{2 \omega } \!\!\!&=&\!\!\! \displaystyle 11
\left(\frac{n_e}{10^{11}~\mbox{cm}^{-3}}\right)
\!\left(\frac{1~\mbox{TeV}}{\omega}\right)\mbox{pc}^{-1}\!,
\end{array}
\]
$\omega$ is the photon (ALP) energy,  $\omega _p^2=4\pi \alpha n_e/m_e$ is
the plasma frequency squared, $n_e$ is the
electron density, $B$ is the magnetic-field component
perpendicular to the beam,
$m_e$ is the electron mass and $\alpha $ is the fine-structure
constant.

The remaining notation in Eq.~(\ref{eq:conversion}) is
\[
\Delta_{{\rm Q},\parallel(\perp)}=\frac{m^{2}_{\gamma
,\parallel(\perp)}}{2\omega},
\]
where $m^{2}_{\gamma ,\parallel(\perp)}$ is the effective mass square of
the longuitudinal (transverse) photon due to interaction with
the external magnetic field in QED.
The critical parameter here is
\[
\kappa=\frac{1}{m_{e}^{3}} \sqrt{(eF_{\mu\nu}l_{\nu})^{2}}=
\frac{\omega}{m_{e}} \, \frac{B_{\perp}}{B_{\rm cr}}
\]
\[
\approx
4.4 \times 10^{-8} \left(\frac{\omega}{1~\mbox{TeV}}\right)
\left(\frac{B}{1~{\rm G}} \right),
%\label{Eq:kappa}
\]
where $l_{\nu }$ is the photon 4-momentum
and $B_{\rm cr}=m_{e}^{2}/e\approx 4.4 \times 10^{13}$~G.
For $\kappa \ll 1$,
\[
\Delta_{Q,\perp}=-1.43 \times 10^{4}
\left(\frac{\omega}{1~{\rm TeV}} \right)
\left(\frac{B}{1~{\rm G}} \right)^{2}
~{\rm pc}^{-1},
\]
\[
\Delta_{Q,\parallel}=\frac{4}{7}\Delta_{Q,\perp}\]
(see Ref.~\cite{FRT} for general $\kappa$).

The strong mixing happens whenever
%\begin{equation}
\[
4\Delta_M^2                                       \gg
\left(\Delta_p+\Delta_{Q,\perp}-\Delta_m \right)^2,
%\label{Eq:strong-mixing}
\]
%\end{equation}
which, banning fine-tuned cancellations, means $\Delta_{m} \ll 2
\Delta_{M}$, $\Delta_{p} \ll 2 \Delta_{M}$ and $|\Delta_{Q,\perp}| \ll 2
\Delta_{M}$, that is
%\begin{equation}
\[
\omega \gg
2.55~\mbox{MeV}
\left(\frac{m}{10^{-7}~\mbox{eV}} \right)^2
\left(\frac{B}{\mbox{G}} \right)^{-1}
\left(\frac{M}{10^{10}~\mbox{GeV}} \right),
%\label{eq:max-mix-m}
%\end{equation}
\]
\[
%\begin{equation}
n_e \ll
2.8 \times 10^{12}~\mbox{cm}^{-3}
\left(\frac{\omega }{1~\mbox{TeV}} \right)
\left(\frac{B}{\mbox{G}} \right)
\left(\frac{M}{10^{10}~\mbox{GeV}} \right)^{-1},
%\label{eq:max-mix-p}
%\end{equation}
\]
\[
%\begin{equation}
\omega \ll
11~\mbox{GeV}
\left(\frac{B}{\mbox{G}} \right)^{-1}
\left(\frac{M}{10^{10}~\mbox{GeV}} \right)^{-1}
%\label{eq:max-mix-q}
%\end{equation}
\]
(the latter inequality assumes $\kappa \ll 1$).
In addition,
the size $L$ of
the magnetic-field region
should exceed the oscillation length,
$
L \gtrsim \frac{\pi}{\Delta_{\rm osc}},
$
which translates into
\begin{equation}
L \gtrsim
0.01~\mbox{pc}
\left(\frac{B}{\mbox{G}} \right)^{-1}
\left(\frac{M}{10^{10}~\mbox{GeV}} \right).
\label{eq:max-mix-losc}
\end{equation}

\section{Parameters: two scenarios}
Two particular scenarios involving ALPs have been proposed
to reduce the opacity of the Universe for
TeV gamma rays from blazars.

The first one implies intergalactic magnetic fields strong
enough to satisfy conditions  for efficient ALP/photon mixing all along
the path between the source and the observer. This mechanism was first
proposed in Ref.~\cite{Csaba} in a different context
and invoked for the TeV blazar spectra in
Ref.~\cite{DARMA}. The photon/ALP mixed beam
propagates through the Universe in this case and, since the photons are
attenuated while ALPs are not, the effective suppression of the flux is
smaller compared to the pure-photon case.
One can easily derive that, for
propagation through domains of randomly oriented magnetic fields, the
optical depth is effectively reduced by 2/3  in the
long-distance limit. A more detailed recent study
\cite{Galanti-spindex} results in the following constraints on
the relevant ALP parameters:  $m \lesssim
10^{-9}$~eV and the ALP-photon coupling $g_{a\gamma \gamma}$ is determined
from $\xi \equiv (B/{\rm nG}) (g_{a\gamma \gamma} \times 10^{11}~{\rm
GeV}) \gtrsim 0.3$, that is $g_{a\gamma \gamma} \gtrsim 3 \times
10^{-12}$~GeV$^{-1}$.

In the frameworks of the second approach,  quite
strong magnetic fields are assumed in the source and
around the observer, while throughout the way,
fields are too weak for ALP/photon mixing. Therefore, the ALP/photon
conversion is efficient in the blazar itself
and in the Milky Way~\cite{Serpico}. Alternatively, it may happen in the
galaxy clusters or filaments surrounding the source and/or the
observer \cite{FRT} (see also a more detailed subsequent study in
Ref.~\cite{1207.0776clusters}).
In this case, a part of emitted photons is converted to ALPs and then
travel intact to the vicinity of the observer, while remaining photons
attenuate in a usual way. In the Milky Way, a part of ALPs convert back to
photons, which are observed. In this way, the flux suppression (i.e., the
effective opacity) does not depend on the distance to the source for large
distances.
A detailed study of this mechanism is
presented, e.g., in Ref.~\cite{Meyer-evidence} (it is called ``the
general-source'' scenario there). Note that
$g_{a\gamma \gamma} \gtrsim 2 \times 10^{-11}$~GeV$^{-1}$
is required in this scenario because for
lower values of the coupling, the path of the ALP-photon beam through the
magnetic field would be too short for efficient conversion even in the
maximal-mixing case.

\begin{figure}
\centerline{\includegraphics[width=0.95\columnwidth]{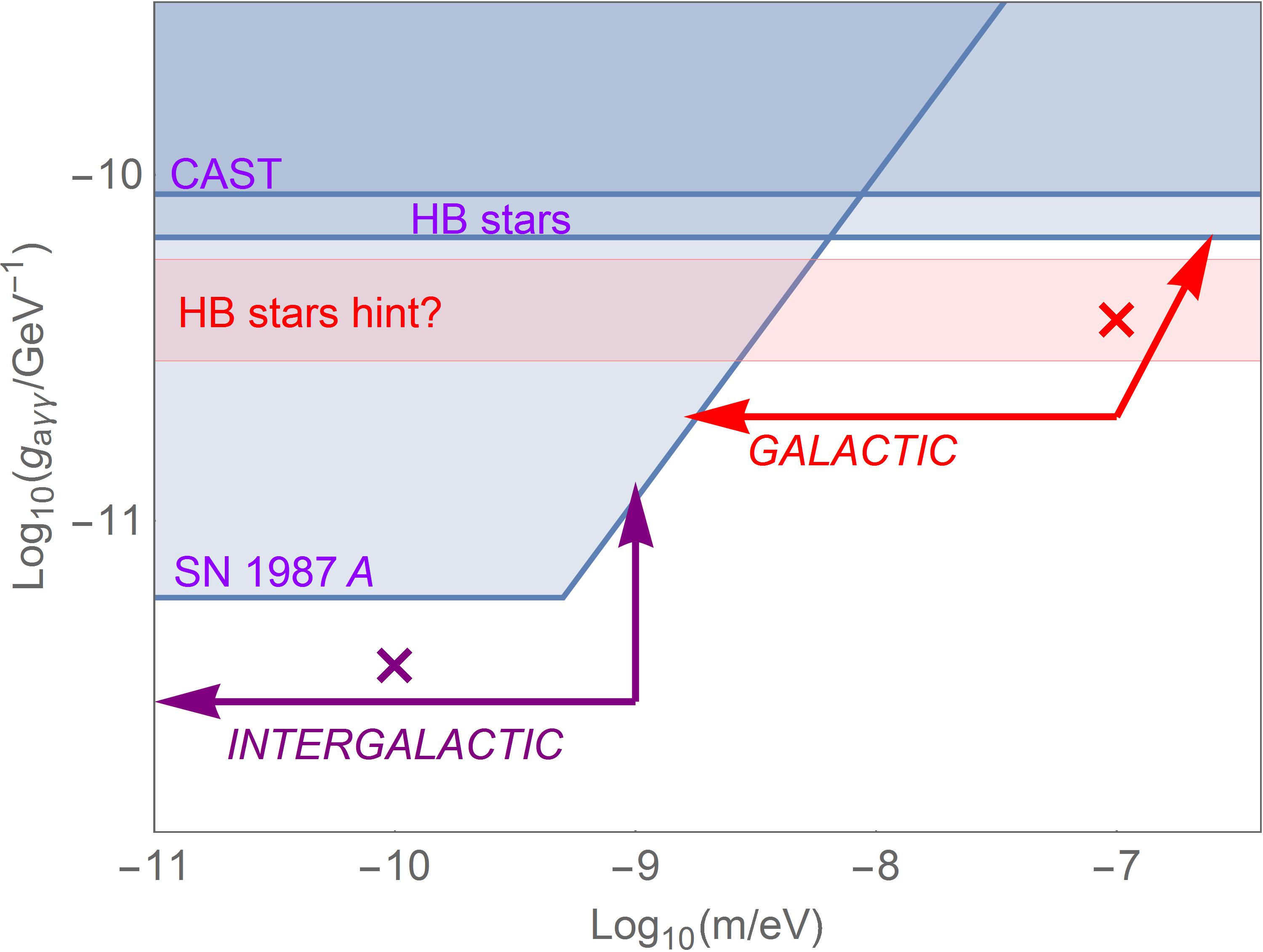}}
\caption{\label{fig:main}
Figure~\ref{fig:main}.
The ALP parameter space (ALP-photon coupling $g_{a\gamma
\gamma}$ versus ALP mass $m$). Blue shade indicates present
constraints (the CERN axion
solar telescope (CAST, Ref.~\cite{CAST}), evolution of the
horizonthal-branch (HB) stars~\cite{HBstars} and reanalysis of the
SN~1987A data \cite{SN1987Anew}.
The pink shaded band is favoured by the horizontal-branch star
cooling~\cite{HBstars}. Regions corresponding to the
galactic~\cite{Galanti-spindex} and intergalactic~\cite{Meyer-evidence}
ALP/photon conversion scenarios are shown. Crosses denote benchmark
parameter values discussed in the text. }
\end{figure}

Figure~\ref{fig:main}
presents the ALP parameter space with present experimental and
observational constraints. Regions relevant for the two scenarios
discussed above are denoted as ``galactic'' and ``intergalactic'',
respectively.
Benchmark parameter values for both scenarios are shown by crosses in
Fig.~\ref{fig:main} ($m=10^{-7}$~eV, $g_{a \gamma
\gamma}=10^{-10.4}$~GeV$^{-1}$ and  $m=10^{-10}$~eV, $g_{a \gamma
\gamma}=10^{-11.4}$~GeV$^{-1}$)\footnote{For the ``galactic'' benchmark,
the photon/ALP conversion also explains~\cite{FRT} puzzling
BL~Lac/cosmic ray correlations observed in HiRes data \cite{BL1, BL2}.}.
Note that for the ``galactic'' scenario, $g_{a \gamma \gamma
} $ fits well the explanation of anomalous cooling of horizontal-branch
stars~\cite{HBstars}, $g_{a\gamma \gamma }=\left( 0.45^{+0.12}_{-0.16}
\right)\times 10^{-10}$~GeV$^{-1}$, see e.g.\ Ref.~\cite{1512.08108} for a
recent review.

Presently, strict upper bounds on the intergalactic magnetic fields
\cite{IGMF} together with the SN1987A constraints~\cite{SN1987Anew} start
to disfavour the intergalactic-mixing scenario, while some weak
evidence for the Galactic anisotropy in the distribution of distant
gamma-ray blazars over the sky~\cite{Serpico, ST-2-scenarios} and
persistence of the pair-production anomaly up to high
redshifts~\cite{gamma} start to favour the Galactic-mixing model. Future
astrophysical studies will be able to distinguish between the scenarios
\cite{ST-2-scenarios}.

One more approach to the search for ALP signatures, or to constraining ALP
parameters, is based on the expected turbulence of cosmic magnetic fields,
which under certain conditions may result in dips and bumps in the spectra
of gamma-ray sources residing, e.g., in galaxy clusters. Two
studies, based on HESS~\cite{HESSirreg} and FERMI-LAT \cite{FERMIirreg}
observations of powerful gamma-ray emitting sources in clusters reported
non-observation of these spectral irregularities and therefore claimed the
exclusion of certain regions in the ALP parameter space, overlapping with
those relevant for the Galactic scenario. However, the magnetic fields in
the corresponding clusters were never measured, and the exclusions are
based on assumed magnetic-field models. Addition, for instance, of a
regular magnetic-field component, which one might expect to present in the
vicinity of a powerful active galaxy, would change the results completely.
A safer way to search for the spectral irregularities is to use
Galactic sources and to invoke models of the Galactic magnetic field
which, though uncertain, are at least based on observations. Preliminary
results of a study of this kind, reported in Ref.~\cite{HornsPoster},
might even favour the existence of an ALP with parameters relevant for the
Galactic mixing scenario.

\section{Future}
\label{sec:future}
Future gamma-ray observations will help to verify the existence of the
anomalous transparency of the Universe for gamma rays and to further
narrow the parameter space of relevant ALPs. They will include, in
particular:
\begin{itemize}
 \item[(i)] high-statistics measurements of spectral energy distributions
of relatively nearby blazars at the energies where the expected opacity is
large, that is at energies $\sim (50-100)$~TeV;
\item[(ii)]
observation of very distant sources, for which the absorption becomes
important at $\sim (10-100)$~GeV;
\item[(iii)]
complete full-sky surveys of distant gamma-ray sources which may reveal or
exclude clear patterns of the Galactic anisotropy.
\end{itemize}
The goal (i) may be achieved with extensive-air-shower arrays equipped
with muon detectors, notably Carpet-2$+$ \cite{Carpet}, TAIGA \cite{TAIGA}
and LHAASO \cite{LHAASO}. To reach the goal (i), one would need to employ
low-threshold high-altitude Cerenkov telescopes \cite{low-threshold};
presently, the project of the Atmospheric Low-Energy Gamma-Ray Observatory
(ALEGRO) is proposed which may reside either in Atacama, Chile, or at the
Mount Elbrus, Russia (within the Elbrus Gamma-ray Observatory, EGO). The
goal (iii) may be realized with future survey-mode data of the Cerenkov
Telescope Array (CTA) \cite{CTA}, but the use of the FERMI-LAT data may
help as well.

The Any Light Particle Search experiment  in its
proposed upgraded configuration (ALPS-II) \cite{ALPS-II} may probe the
strongest values of $g_{a\gamma \gamma }$ relevant for the Galactic
mixing scenario. Both the Galactic and intergalactic scenarios correspond
to the discovery region of the International Axion Observatory (IAXO)
\cite{NGAH, IAXO_CDR}. For the slightly favoured Galactic case, the
number of events in IAXO would allow for a detailed study of the ALP by
means of change of the magnetic-field geometry. In fact, even a smaller
instrument would be able to discover the ALP in this case.

\section*{Acknowledgements}
The author is indebted to M.~Giannotti, I.~Irastorza and M.~Meyer for
interesting discussions. This
work is supported by the Russian Science Foundation, grant 14-12-01340.

\end{document}